\documentclass[dvips,12pt,a4paper]{article}
\usepackage{a4wide}
\usepackage{epsfig}
\usepackage{amsmath}
\usepackage{latexsym}

  \newlength{\absize}
\newcommand{\dd}{\mbox{{\rm d}}}

\newcommand{\Lumint}{{\cal L}_{\rm int}}

\allowdisplaybreaks %!!!!

\catcode`@=11
\def\citer{\@ifnextchar [{\@tempswatrue\@citexr}{\@tempswafalse\@citexr[]}}
 
\def\@citexr[#1]#2{\if@filesw\immediate\write\@auxout{\string\citation{#2}}\fi
  \def\@citea{}\@cite{\@for\@citeb:=#2\do
    {\@citea\def\@citea{--\penalty\@m}\@ifundefined
       {b@\@citeb}{{\bf ?}\@warning
       {Citation `\@citeb' on page \thepage \space undefined}}%
\hbox{\csname b@\@citeb\endcsname}}}{#1}}
\catcode`@=12

\begin{document}
  \thispagestyle{empty}
  \pagestyle{empty}
  \renewcommand{\thefootnote}{\fnsymbol{footnote}}
\newpage\normalsize
    \pagestyle{plain}
    \setlength{\baselineskip}{4ex}\par
    \setcounter{footnote}{0}
    \renewcommand{\thefootnote}{\arabic{footnote}}
\newcommand{\preprint}[1]{%
  \begin{flushright}
    \setlength{\baselineskip}{3ex} #1
  \end{flushright}}
\renewcommand{\title}[1]{%
  \begin{center}
    \LARGE #1
  \end{center}\par}
\renewcommand{\author}[1]{%
  \vspace{2ex}
  {\Large
   \begin{center}
     \setlength{\baselineskip}{3ex} #1 \par
   \end{center}}}
\renewcommand{\thanks}[1]{\footnote{#1}}
\renewcommand{\abstract}[1]{%
  \vspace{2ex}
  \normalsize
  \begin{center}
    \centerline{\bf Abstract}\par
    \vspace{2ex}
    \parbox{\absize}{#1\setlength{\baselineskip}{2.5ex}\par}
  \end{center}}

\begin{flushright}
{\setlength{\baselineskip}{2ex}\par
{\tt University of Bergen, Department of Physics}    \\[1mm]
{\tt Scientific/Technical Report No.\ 2001-01}    \\[1mm]
{\tt ISSN 0803-2696} \\[2mm]
{hep-ph/0101150} \\[2mm]
{January 2001}           \\
} %end tighter \baselineskip
\end{flushright}
\vspace*{4mm}
\vfill
\title{Contact interactions and polarized beams \\
at a Linear Collider}
\vfill
\author{
A.A. Babich$^{a}$,
P. Osland$^{b}$,
A.A. Pankov$^{a,c}$ {\rm and}
N. Paver$^{c}$}
%-----------------------------------
%   Address
%-----------------------------------
%\vspace{1cm}
\begin{center}
$^a$ Pavel Sukhoi Technical University, 
     Gomel, 246746 Belarus \\
$^b$ Department of Physics, University of Bergen, \\
     All\'{e}gaten 55, N-5007 Bergen, Norway \\
$^c$ Dipartimento di Fisica Teorica, Universit\`a di Trieste and \\
Istituto Nazionale di Fisica Nucleare, Sezione di Trieste,
Trieste, Italy
\end{center}
\vfill
\abstract
{We discuss contact-interaction searches in the processes 
$e^+e^-\to \mu^+\mu^-$, $b\bar{b}$ and $c\bar{c}$ at an  
$e^+e^-$ Linear Collider with c.m.\ energy $\sqrt{s}=0.5$ TeV and 
with longitudinally polarized beams. The measurement of polarized cross 
sections allows to study the individual helicity cross sections, and 
consequently to derive separate, model-independent, constraints on the 
four-fermion contact interaction couplings. We evaluate the reach on those
parameters foreseeable in the case of both electron and positron polarization 
fixed at some reference values, and compare it with the situation where only 
electron polarization is available. 
The analysis is based on polarized 
integrated cross sections with optimal kinematical cuts that can improve the 
sensitivity to the relevant couplings. 
While electron polarization would by itself allow such an analysis, 
the additional 
positron polarization (with no loss of beam intensity) and optimization
can have a crucial role in improving the sensitivity to the new
interactions.}
\vspace*{20mm}
\setcounter{footnote}{0}
\vfill

\newpage
    \setcounter{footnote}{0}
    \renewcommand{\thefootnote}{\arabic{footnote}}
    \setcounter{page}{1}

%%%%%%%%%%%%%%%%%%%%%%%%%%%%%%%%%%%%%%%%%%%%%%%%%%%%%%%%%%%%%%%%%%%%%%%%
\section{Introduction}
%%%%%%%%%%%%%%%%%%%%%%%%%%%%%%%%%%%%%%%%%%%%%%%%%%%%%%%%%%%%%%%%%%%%%%%%

The possibility of longitudinally polarizing electron and positron beams 
at the Linear Collider (LC) is considered with great interest in connection 
with the physics programme to be performed at such a facility 
\cite{Accomando}. 
Indeed, this situation would enable to probe with enhanced sensitivity 
the chiral structure of electroweak interactions and, in particular, to 
set stringent, and model-independent, constraints on new interactions 
by looking for deviations of the data from the Standard Model (SM) 
predictions. 
\par 
Here, we will consider the process of fermion pair production 
($f\ne e$, $t$)
\begin{equation}
e^++e^-\to f+\bar{f} \label{proc}
\end{equation}
%\rightline{proc}
at the LC with: (i) one beam (electron) polarized, and (ii) both beams 
polarized. For both cases, and for different values of the luminosity, we 
discuss the sensitivity of the measurable 
helicity cross sections to the $SU(3)\times SU(2)\times U(1)$ 
symmetric $eeff$ contact-interaction effective Lagrangian with 
helicity-conserving and flavor-diagonal fermion currents \cite{Eichten}:       
\begin{equation}
{\cal L}_{\rm C.I.}
=\sum_{\alpha\beta}\frac{g^2_{\rm eff}}{\Lambda^2_{\alpha\beta}}
\eta_{\alpha\beta}\left(\bar e_{\alpha}\gamma_\mu e_{\alpha}\right)
\left(\bar f_{\beta}\gamma^\mu f_{\beta}\right).
\label{lagra}
\end{equation}
%\rightline{lagra}
In Eq.~(\ref{lagra}), generation and color indices have been suppressed,
$\alpha,\beta={\rm L,R}$ indicate left- or right-handed helicities, and 
$\eta_{\alpha\beta}=\pm 1,0$ depending on the chiral structure of the
individual interactions. Also, one takes $g^2_{\rm eff}=4\pi$ to remind that 
such new interaction, originally proposed for compositeness, would become 
strong at $\sqrt s\sim\Lambda_{\alpha\beta}$. However, more generally 
${\cal L}$ should be considered as the `low-energy' parameterization 
of some non-standard interaction acting at the much larger energy scales 
$\Lambda$ not attainable by the machine. Examples are the exchanges in the 
different channels of extremely heavy objects such as $Z^\prime$ with 
a few TeV mass \cite{barger} and leptoquarks \cite{altarelli}. 
\par 
Clearly, the new coupling constants in Eq.~(\ref{lagra}) (equivalently, the 
mass scales $\Lambda_{\alpha\beta}$) are {\it a priori} free parameters that 
induce deviations of observables from the SM predictions, and the attainable 
constraints are assessed by the numerical comparison of such deviations to the 
expected experimental accuracies. 
\par 
For a given final fermion species $f$, Eq.~(\ref{lagra}) defines eight 
individual, independent models corresponding to the combinations of the four 
chiralities $\alpha,\beta$ with the $\pm$ signs of the $\eta$'s. Therefore, in 
the most general case where the observed contact interaction is a combination 
of these models, one faces the complication that the 
aforementioned deviations simultaneously depend on all four-fermion effective 
couplings in 
Eq.~(\ref{lagra}). The simplest procedure consists in assuming a non-zero 
value for only one coupling at a time (or a specific combination of them)
and a 1-parameter $\chi^2$ fit to the
data, which leads to tests of the models mentioned above  
\cite{barger,Riemann-00}. 
\par
On the other hand, a general, model-independent, analysis 
must simultaneously account for all non-zero couplings as free parameters 
and, at the same time, allow the derivation of separate constraints. This 
possibility is offered by initial beam polarization, that enables the
extraction from the data of the individual helicity cross sections of process 
(\ref{proc}), each one being directly related to a single $eeff$ contact term 
that, accordingly, can be disentangled.  
\par
Actually, we shall adopt here as basic observables two particular, 
polarized, integrated cross sections that allow to reconstruct the four 
helicity amplitudes from linear combinations of measurements at different 
values of the beam polarizations. Integrated observables
should be of some advantage in the case of limited experimental statistics. 
Also, in principle, a significant improvement can be obtained 
by defining optimally chosen kinematical regions of integration that  
lead to maximal sensitivity of the analysis to the four-fermion 
couplings \cite{BOPP-99,BOPP-00}. 
\par 
In the sequel, after giving the main definitions and briefly reviewing the 
procedure, we shall assess the reach on $\Lambda_{\alpha\beta}$  
for the LC with $\sqrt s=0.5\ {\rm TeV}$ as a function of the luminosity, 
for reference values of the electron and positron longitudinal polarizations 
with given uncertainties, and making standard assumptions on the expected 
systematic uncertainties on the cross sections for process (\ref{proc}). 

%%%%%%%%%%%%%%%%%%%%%%%%%%%%%%%%%%%%%%%%%%%%%%%%%%%%%%%%%%%%%%%%%%%%%%%%
\section{Determination of helicity cross sections}
%%%%%%%%%%%%%%%%%%%%%%%%%%%%%%%%%%%%%%%%%%%%%%%%%%%%%%%%%%%%%%%%%%%%%%%%
Limiting ourselves to the cases $f\neq e, t$ and neglecting the fermion 
mass with respect 
to the c.m. energy $\sqrt s$, the differential cross section of process 
(\ref{proc}) with polarized electron and positron beams reads, in the Born
approximation \cite{Zeppenfeld2}: 
\begin{equation}
\frac{\dd\sigma}{\dd\cos\theta}
=\frac{3}{8}
\left[(1+\cos\theta)^2 {\sigma}_+
+(1-\cos\theta)^2 {\sigma}_-\right].
\label{cross}
\end{equation}                       
Here, $\theta$ is the angle between the incoming electron and the outgoing 
fermion in the c.m. frame and, with $P_e$ and $P_{\bar e}$ the longitudinal 
polarizations of the beams, $\sigma_+$ and $\sigma_-$ can be expressed in 
terms of the helicity cross sections as
\begin{eqnarray}
{\sigma}_{+}&=&\frac{1}{4}\,
\left[(1-P_e)(1+P_{\bar{e}})\,\sigma_{\rm LL}
+(1+P_e)(1- P_{\bar{e}})\,\sigma_{\rm RR}\right]\nonumber \\
&=&\frac{D}{4}\,\left[(1-P_{\rm eff})\,\sigma_{\rm LL}
+(1+P_{\rm eff})\,\sigma_{\rm RR}\right], 
\label{s+} \\
{\sigma}_{-}&=&\frac{1}{4}\,
\left[(1-P_e)(1+ P_{\bar{e}})\,\sigma_{\rm LR}
+(1+P_e)(1-P_{\bar{e}})\,\sigma_{\rm RL}\right] \nonumber \\
&=&
\frac{D}{4}\,\left[(1-P_{\rm eff})\,\sigma_{\rm LR}
+(1+P_{\rm eff})\,\sigma_{\rm RL}\right], \label{s-}
\end{eqnarray}
where 
\begin{equation}
P_{\rm eff}=\frac{P_e-P_{\bar{e}}}{1-P_eP_{\bar{e}}} 
\label{pg}
\end{equation}
is the effective polarization \cite{Flottmann-Omori}, satisfying
$\vert P_{\rm eff}\vert\leq 1$, and $D=1-P_eP_{\bar{e}}$.
%\rightline{cross}
It should be noted that with $P_{\bar{e}}\ne0$, $\vert P_{\rm eff}\vert$
can be larger than $|P_e|$.
Moreover, with $\alpha,\beta={\rm L,R}$, in Eqs.~(\ref{s+}) and (\ref{s-}):  
\begin{equation}
\sigma_{\alpha\beta}=N_C\sigma_{\rm pt}
\vert A_{\alpha\beta}\vert^2,
\label{helcross}
\end{equation}
%\rightline{helcross}
where $N_C\approx 3(1+\alpha_s/\pi)$ for quarks and $N_C=1$ for leptons, 
respectively, and $\sigma_{\rm pt}\equiv\sigma(e^+e^-\to\gamma^\ast\to l^+l^-)
=(4\pi\alpha^2)/(3s)$. Including the $\gamma,\ Z$ exchanges and the 
contribution of ${\cal L}_{\rm C.I.}$ according to Eq.~(\ref{lagra}), 
the helicity amplitudes $A_{\alpha\beta}$ can be written as
\begin{equation}
A_{\alpha\beta}=Q_e Q_f+g_\alpha^e\,g_\beta^f\,\chi_Z+
\frac{s\eta_{\alpha\beta}}{\alpha\Lambda_{\alpha\beta}^2},
\label{amplit}
\end{equation}
%\rightline{amplit}
where $\chi_Z=s/(s-M^2_Z+iM_Z\Gamma_Z)$ is the $Z$ boson propagator,
$g_{\rm L,R}^f$ are the SM left- and right-handed fermion 
couplings of the $Z$, and $Q_f$ are the fermion electric charges.
The above relations clearly show the direct relation of helicity cross 
sections to the individual contact interactions in Eq.~(\ref{lagra})
with definite chiralities, that allows the desired model-independent analysis 
with all contact-interactions taken into account simultaneously as free 
parameters. The various contributions in Eqs.~(\ref{s+}) and (\ref{s-}) 
can be disentangled by making measurements at two different values of 
the polarizations (a minimum of four measurements is needed). For this 
purpose we use the set of values $P_e=\pm P_1$ and $P_{\bar e}=\mp P_2$ 
($P_{1,2}>0$) or, alternatively, $P_{\rm eff}=\pm P$ with $D$ fixed. 
Correspondingly, from Eqs.~(\ref{s+}) and (\ref{s-}):
\begin{eqnarray}
\label{SLL}
\sigma_{\rm LL}
&=&\frac{1}{D}\left[
- \frac{1-P}{P}\sigma_{+}(P)
+\frac{1+P}{P}\sigma_{+}(-P) \right], \\
\label{SRR}
\sigma_{\rm RR}
&=&\frac{1}{D}\left[\frac{1+P}{P}\sigma_{+}(P) 
- \frac{1-P}{P}\sigma_{+}(-P)\right],
\end{eqnarray} 
with $\sigma_{\rm LR}$ and $\sigma_{\rm RL}$ obtained from 
$\sigma_{\rm LL}$ and $\sigma_{\rm RR}$, respectively, 
replacing $\sigma_{+}$ by $\sigma_{-}$.

Actually, for the purpose of optimizing the resulting 
bounds on $\Lambda_{\alpha\beta}$, one can more generally define the polarized
cross sections integrated over the {\it a priori} arbitrary kinematical ranges
($-1,\ z^*$) and ($z^*,\ 1$) \cite{BOPP-99}: 
\begin{eqnarray}
\label{sigma1}
\sigma_1(z^*, P, D)
&\equiv&\int_{z^*}^1\frac{\dd\sigma}{\dd\cos\theta}\dd\cos\theta
=\frac{1}{8}\left\{\left[8-(1+z^*)^3\right]\sigma_++(1-z^*)^3
\sigma_-\right\}, \\
\label{sigma2}
\sigma_2(z^*, P,{D})
&\equiv&\int^{z^*}_{-1}\frac{\dd\sigma}{\dd\cos\theta}\dd\cos\theta
=\frac{1}{8}\left\{(1+z^*)^3\sigma_++
\left[8-(1-z^*)^3\right]\sigma_-\right\},  
\end{eqnarray}
and take $\sigma_{1,2}(z^*, P, D)$ as the basic set of integrated polarized 
observables to be measured.\footnote{For simplicity of notations, the 
polarization dependence of $\sigma_{\pm}$ on the right-hand sides of 
Eqs.~(\ref{sigma1}) and (\ref{sigma2}) has been suppressed.} 
The basic reason this procedure, with $z^*\ne0$, can be advantageous,
is that the SM amplitude, against which the contact  interaction term
interferes, is {\it not} forward-backward symmetric.
By solving Eqs.~(\ref{sigma1}) and (\ref{sigma2}) one obtains $\sigma_+$ and 
$\sigma_-$ from the measurement of $\sigma_1$ and $\sigma_2$: 
\begin{eqnarray}
\label{sigmap}
\sigma_+
&=&\left[a(z^*)\sigma_1(z^*,P,D) 
        +b(z^*)\sigma_2(z^*,P,D)\right], \\
\label{sigmam}
\sigma_-
&=&\left[b(-z^*)\sigma_1(z^*,P,D)
        +a(-z^*)\sigma_2(z^*,P,D)\right],
\end{eqnarray}
where
\begin{equation}
\label{Eq:a-b}
a(z^*)=\frac{8-(1-z^*)^3}{6(1-{z^*}^2)}, \qquad
b(z^*)=-\frac{(1-z^*)^3}{6(1-{z^*}^2)}.
\end{equation}
The experimental values of the helicity cross sections 
$\sigma_{\alpha\beta}$ are finally determined from the linear system of 
equations (\ref{SLL}), (\ref{SRR}). 
\par 
The advantage of this, rather elaborate, procedure is that the actual 
value of $z^*$, representing an input parameter related to given experimental 
conditions, can be tuned to achieve an optimization
of the constraints on the mass scales $\Lambda_{\alpha\beta}$.
\par 
Of course, electron {\it or} positron polarization is a 
{\it necessity} in order to disentangle the helicity cross sections and 
evaluate separate, and model-independent, constraints on the corresponding 
contact-interaction couplings. However, one can expect on statistical 
grounds a significant increase of the sensitivity due to the polarization of 
positrons, provided the luminosity in this case remains the same or is 
only moderately reduced, and the polarization is known very precisely.     

In the following analysis, cross sections will be evaluated including 
initial- and final-state radiation by means of the program 
ZFITTER \cite{zfitter}, which has to be used along with ZEFIT, adapted to 
the present discussion, with $m_{\rm top}=175$~GeV and
$m_H=120$~GeV. One-loop SM electroweak corrections are accounted for by 
improved Born amplitudes \cite{Hollik,Altarelli2}, such that the form of the 
previous formulae remains the same. Concerning initial-state radiation, a cut 
on the energy of the emitted photon $\Delta=E_\gamma/E_{\rm beam}=0.9$ is 
applied for $\sqrt s=0.5\ {\rm TeV}$ in order to avoid the radiative return 
to the $Z$ peak, and increase the signal originating from the contact 
interaction contribution \cite{Djouadi}.  

%%%%%%%%%%%%%%%%%%%%%%%%%%%%%%%%%%%%%%%%%%%%%%%%%%%%%%%%%%%%%%%%%%%%%%%%
\section{Sensitivity of polarized observables}
%%%%%%%%%%%%%%%%%%%%%%%%%%%%%%%%%%%%%%%%%%%%%%%%%%%%%%%%%%%%%%%%%%%%%%%%

Current bounds on $\Lambda_{\alpha\beta}$, of the order of several 
TeV \cite{Lambda}, are such that for the LC c.m.\ energy $\sqrt{s}=0.5$ TeV 
the characteristic suppression factor $s/\Lambda^2$ in Eq.~(\ref{amplit}) 
is quite small. Therefore, we can only look at indirect manifestations of the 
contact interaction (\ref{lagra}) as deviations of measured helicity cross
sections from the SM predictions, and assess the corresponding reach on the 
$\Lambda_{\alpha\beta}$ on the basis of the foreseen initial beam
polarizations and the experimental accuracies.     
\par
We can define the `sensitivity' to contact interactions of each helicity 
cross section as the ratio
\begin{equation}
\label{signif}
{\cal S}(\sigma_{\alpha\beta})
=\frac{|\Delta\sigma_{\alpha\beta}|}{\delta\sigma_{\alpha\beta}}, 
\end{equation}
%\rightline{signif}
where $\Delta\sigma_{\alpha\beta}$ is the deviation from the SM prediction 
due to (\ref{lagra}), dominated for $\sqrt s\ll \Lambda_{\alpha\beta}$ by the 
linear interference term 
\begin{equation}
\Delta\sigma_{\alpha\beta}\equiv
\sigma_{\alpha\beta}-\sigma_{\alpha\beta}^{\rm SM}\simeq
2 N_C\, \sigma_{\rm pt}
\left(Q_e\, Q_f+g_{\alpha}^e\, g_{\beta}^f\,\chi_Z\right)
\frac{s\eta_{\alpha\beta}}{\alpha\Lambda_{\alpha\beta}^2},
\label{deltasig}
\end{equation}
and $\delta\sigma_{\alpha\beta}$ denotes the expected experimental uncertainty
on $\sigma_{\alpha\beta}$, combining statistical and systematic uncertainties.
The reach on $\Lambda_{\alpha\beta}$ can be obtained from a $\chi^2$ 
analysis,
\begin{equation}
\label{chisq}
\chi^2\equiv{\cal S}^2
=\left(\frac{\Delta\sigma_{\alpha\beta}}{\delta\sigma_{\alpha\beta}}\right)^2,
\end{equation}
by imposing, as a criterion to constrain the allowed values of the 
contact-interaction parameters from the non-observation of the corresponding 
deviations within the expected uncertainty $\delta\sigma_{\alpha\beta}$, 
that: 
\begin{equation}
\chi^2<\chi^2_{\rm CL},
\end{equation}
where the actual value of $\chi^2_{\rm CL}$ specifies the desired 
`confidence' level. Since, as (\ref{deltasig}) shows, the deviation
$\Delta\sigma_{\alpha\beta}$ depends on a single `effective' non-standard 
parameter represented by the product of the known relevant SM coupling 
times the contact-interaction coupling one wants to constrain, in such 
a $\chi^2$ analysis of data one effective parameter is involved, and we take 
$\chi^2_{\rm CL}=3.84$ corresponding to 95\% C.L. with a one-parameter 
fit.\footnote{The signs of the $\eta$'s in (\ref{lagra}) turn out to be 
numerically unimportant for the determination of constraints on the
$\Lambda_{\alpha\beta}$. Indeed, for given helicities $\alpha\beta$, 
different signs of $\eta$'s yield practically identical results for the mass 
scales $\Lambda_{\alpha\beta}$ as long as, in the chosen kinematical 
configurations, the non-standard effects are largely dominated by the 
interference (\ref{deltasig}) between contact-interaction and SM terms.} 
\par 
To proceed to numerical evaluations of the bounds, an assessment of the 
expected experimental uncertainty $\delta\sigma_{\alpha\beta}$ is needed. 
To obtain an indication,
we combine all uncertainties in quadrature, and separate for convenience 
the systematic 
uncertainty of the initial positron and electron polarizations, essentially 
by considering $\sigma_{1,2}$, $P_e$ and $P_{\bar e}$ in 
Eqs.~(\ref{SLL}) and (\ref{SRR}) as if they were independent measurables.
This is clearly an approximation, but it should exhibit the main dependence
on the uncertainties of the polarizations. Thus,
\begin{equation} 
\left(\delta\sigma_{\alpha\beta}\right)^2=
\left({\overline\delta}\sigma_{\alpha\beta}\right)^2+
\left(\delta\sigma^{\rm pol}_{\alpha\beta}\right)^2.
\label{delpol}
\end{equation} 
With $\sigma_{1,2}$ our basic observables (see
Eqs.~(\ref{SLL})--(\ref{Eq:a-b})), one can write: 
\begin{eqnarray}
\label{uncet1}
\left({\overline\delta}\sigma_{\rm LL}\right)^2
&=&a^2(z^*)\left[
\left(\frac{1-P}{P D}\right)^2 
(\delta\sigma_1 (z^*,P))^2
+
\left(\frac{1+P}{P D}\right )^2 
(\delta\sigma_1 (z^*,-P))^2
\right]
\nonumber \\
&+&b^2(z^*)\left[ 
\left ( \frac{1-P}{P D} \right )^2 
(\delta\sigma_2 (z^*,P))^2
+
\left ( \frac{1+P}{P D} \right )^2 
(\delta\sigma_2 (z^*,-P))^2
\right], 
\end{eqnarray}
\begin{eqnarray}
\label{uncet2}
\left({\overline\delta}\sigma_{\rm LR}\right)^2
&=&b^2(-z^*)\left[
\left(\frac{1-P}{P D} \right )^2 
(\delta\sigma_1(z^*,P))^2
+
\left (\frac{1+P}{P D}\right )^2 
(\delta\sigma_1 (z^*,-P))^2 
\right]
\nonumber \\
&+&a^2(-z^*) \left[
\left(\frac{1-P}{P D} \right )^2 
(\delta\sigma_2(z^*,P))^2
+
\left(\frac{1+P}{P D}\right)^2 
(\delta\sigma_2(z^*,-P))^2
\right], 
\end{eqnarray}
where $P=|P_{\rm eff}|$.
Explicit expressions for ${\overline\delta}\sigma_{\rm RR}$ and 
${\overline\delta}\sigma_{\rm RL}$ can be derived from 
${\overline\delta}\sigma_{\rm LL}$ and ${\overline\delta}\sigma_{\rm LR}$, 
respectively, replacing in the above equations 
$\pm P\to \mp P$ in $\delta\sigma_i (z^*,\pm P)$, but not in the corresponding
prefactors. For simplicity of notations, the dependence of 
$\delta\sigma_{1,2}$ on $D$ has not been explicitly indicated.
\par 
The expected smallness of deviations from the SM allows the use, to a very 
good approximation, of the SM predictions for the cross sections 
$\sigma_{1,2}$ to assess the expected $\delta\sigma_1$ and 
$\delta\sigma_2$ in (\ref{uncet1}), (\ref{uncet2}) and, accordingly, to write: 
\begin{equation}
\label{delsi1}
(\delta\sigma_i)^2\simeq(\delta\sigma_i^{\rm SM})^2
=\frac{\sigma_i^{\rm SM}}{\epsilon\, \Lumint}
+\left(\delta^{\rm sys}\sigma_i^{\rm SM}\right)^2, \qquad i=1,2.
\end{equation}
%\rightline{delsi1}
In Eq.~(\ref{delsi1}), $\Lumint$ is the integrated luminosity, and 
$\epsilon$ is the efficiency for detecting the final state under
consideration. For our numerical analysis we shall assume the commonly used 
reference values of the identification efficiencies $\epsilon$ and the 
systematic uncertainties $\delta^{\rm sys}$ \cite{Damerell}: 
$\epsilon=95\%$ and $\delta^{\rm sys}=0.5\%$ for $l^+l^-$; 
$\epsilon=60\%$ and $\delta^{\rm sys}=1\%$ for $b\bar{b}$; $\epsilon=35\%$ 
and $\delta^{\rm sys}=1.5\%$ for $c\bar{c}$. Notice that, as 
a simplification, we take the same $\delta^{\rm sys}$ for both $i=1$ and 2, 
and independent of $z^*$ in the relevant angular range. Concerning the
statistical uncertainty, we shall vary $\Lumint$ from $50$ to 
$500\ \mbox{fb}^{-1}$ (half for each polarization orientation) to 
study the relative roles of statistical and systematic uncertainties, and 
a fiducial experimental angular range $|\cos\theta|\le 0.99$.   
\par 
Let us now turn to a discussion of the systematic uncertainty
of the initial beam polarization.
Finite values of $\delta P_e$ and of $\delta P_{\bar e}$ 
will influence the extraction of the helicity cross sections
$\sigma_{\alpha\beta}$ through the prefactors of 
Eqs.~(\ref{SLL}), (\ref{SRR}), (\ref{sigmap}) and (\ref{sigmam}), as well as 
through the dependence of $\sigma_1$ and $\sigma_2$ on $P$ and $D$. 
Lacking at present sufficiently detailed knowledge of the individual 
sources of uncertainty needed for a complete assessment, 
for simplicity we model the 
systematic uncertainty by assuming the latter effect to be included in  
the $\delta^{\rm sys}\sigma_i$ introduced in Eq.~(\ref{delsi1}).  
Under the above assumptions, we obtain
\begin{eqnarray}
\label{Eq:uncert-P}
\left(\delta \sigma_{\rm LL}^{\rm pol}\right)^2
&=&[f(z^{\ast},P)(1+P_{\bar e}P^2)-f(z^{\ast},-P)(1-P_{\bar e}P^2)]^2
\left(\frac{\delta P_e}{D^2P^2}\right)^2 \nonumber \\
&+&[f(z^{\ast},P)(1-P_{e}P^2)-f(z^{\ast},-P)(1+P_{e}P^2)]^2
\left(\frac{\delta P_{\bar e}}{D^2P^2}\right)^2, \nonumber \\
\left(\delta \sigma_{\rm RR}^{\rm pol}\right)^2
&=&[f(z^{\ast},P)(1-P_{\bar e}P^2)-f(z^{\ast},-P)(1+P_{\bar e}P^2)]^2
\left(\frac{\delta P_e}{D^2P^2}\right)^2 \nonumber \\
&+&[f(z^{\ast},P)(1+P_{e}P^2)-f(z^{\ast},-P)(1-P_{e}P^2)]^2
\left(\frac{\delta P_{\bar e}}{D^2P^2}\right)^2,
\end{eqnarray}
with
\begin{equation}
\label{Eq:f}
f(z^{\ast},P)=a(z^{\ast}) \sigma_1(z^{\ast},P)
+ b(z^{\ast})\sigma_2(z^{\ast},P).
\end{equation}
Furthermore, $\delta \sigma_{\rm LR}^{\rm pol}$ and 
$\delta \sigma_{\rm RL}^{\rm pol}$ are obtained from
$\delta \sigma_{\rm LL}^p$ and $\delta \sigma_{\rm RR}^p$,
respectively, by substituting $a(z^{\ast})\leftrightarrow b(-z^{\ast})$.
Numerically, for explicit evaluations of the reach in $\Lambda_{\alpha\beta}$, 
we shall work out the example of $\vert P_e\vert=0.8$ with 
$\delta P_e/P_e=0.5\%$, as currently achieved \cite{SLC}, 
and $\vert P_{\bar e}\vert=$ 0.0, 0.4 and 0.6 with 
$\delta P_{\bar e}/P_{\bar e}=0.5\%$, 
assuming no loss of luminosity compared to 
the case of no positron polarization. With these values of the longitudinal 
polarizations, $\vert P_{\rm eff}\vert = P =0.8,\ D=1$; 
$P=0.909,\ D=1.32$; $P=0.946,\ D=1.48$, respectively. 
\par 
As far as the proposed optimization procedure is concerned, 
from the previous formulae we observe that the $z^*$ dependence of 
$\sigma_1$ and $\sigma_2$, as defined in Eqs.~(\ref{sigma1}) 
and (\ref{sigma2}), translates into a $z^*$ dependence of the uncertainties 
$\delta\sigma_{\alpha\beta}$ that appear in (\ref{signif}) and 
(\ref{chisq}). Since the deviation $\Delta\sigma_{\alpha\beta}$ 
is independent of $z^*$, see Eq.~(\ref{deltasig}), the full sensitivity of 
each helicity cross section to the relevant contact-interaction coupling 
constant is determined not only by the size but also by the $z^*$ behavior 
of the corresponding uncertainty $\delta\sigma_{\alpha\beta}$. Therefore, 
an optimization would be obtained by choosing for $z^*$ the value 
$z^*_{\rm opt}$ where the uncertainty $\delta\sigma_{\alpha\beta}$ 
has a minimum, {\it i.e.}, where the corresponding sensitivity 
Eq.~(\ref{signif}) has a maximum.  
\par
For the minimization of the statistical uncertainty, the first term 
on the right-hand side of Eq.~(\ref{delsi1}), one may use the explicit 
expressions of the SM cross sections. The equation determining the  
relevant $z^*$ is:
\begin{equation}
z^*=-3 \frac{1-r_{\alpha\beta}}{1+r_{\alpha\beta}}
\frac{{z^*}^4 -6{z^*}^2 -3}{{z^*}^4 -2{z^*}^2 -23},
\label{zopt}
\end{equation}
where
\begin{equation}
\label{rL}
r_{\rm LL}=r_{\rm LR}=\frac{(1+3P_{\rm eff}^2)\sigma_{\rm LR}^{SM}
+(1-P_{\rm eff}^2)\sigma_{\rm RL}^{SM}}
{(1+3P_{\rm eff}^2)\sigma_{\rm LL}^{SM}+(1-P_{\rm eff}^2)\sigma_{\rm RR}^{SM}},
\end{equation}
and $r_{\rm RR}=r_{\rm RL}$ is obtained by replacing 
${\rm L}\leftrightarrow{\rm R}$ in (\ref{rL}). 
As one can see, the location of $z^*$ that
minimizes the statistical uncertainty only depends on the SM parameters
and $P_{\rm eff}$ and for each final-state fermion in (\ref{proc}) 
is independent of the luminosity and the efficiency of reconstruction 
$\epsilon$. 
In a left-right symmetric theory, the above ratios $r_{\alpha\beta}$
would all be 1, and in this case $z^*=0$. However, in the SM, depending on 
flavour and energy, $r_{\alpha\beta}$ may be less than, or larger than unity.
Since the $z^*$-dependent fraction in (\ref{zopt}) is positive
for $z^{*2}\le1$, it follows that the solutions satisfy $z^*<0$
if $r_{\alpha\beta}<1$ and {\it vice versa}.
We also note that the location is the same for the $\rm LL$ and $\rm LR$ 
configurations, and likewise for $\rm RR$ and $\rm RL$, while numerically 
the sensitivities are different. Clearly, the values of $z^*$ 
determined from the above SM formulae can be regarded as a simple, first 
determination of $z^*_{\rm opt}$ in the case where the expected uncertainty 
is dominated by the statistical one ({\it e.g.}, for low luminosities).  
In the cases where statistical and systematic uncertainties are comparable, 
the $z^*_{\rm opt}$ must be determined by a more elaborate numerical 
analysis that includes all different sources of experimental 
uncertainties. A more extended discussion and a set of  
numerical results can be found in Refs.~\cite{BOPP-99, BOPP-00}.
 
%%%%%%%%%%%%%%%%%%%%%%%%%%%%%%%%%%%%%%%%%%%%%%%%%%%%%%%%%%%%%%%%%%%%%%%%
\section{Bounds on \boldmath$\Lambda_{\alpha\beta}$ and concluding remarks}
%%%%%%%%%%%%%%%%%%%%%%%%%%%%%%%%%%%%%%%%%%%%%%%%%%%%%%%%%%%%%%%%%%%%%%%%

We assume half the total integrated luminosity for each value of the 
effective polarization, $P_{\rm eff}=\pm P$, and the same time of operation 
in the different polarization configurations. 
From the procedure and the 
inputs outlined in the previous section, we find for the discovery limits 
on the mass scales $\Lambda_{\alpha\beta}$ {\it vs.} $\Lumint$ the results 
represented by the curves in Fig.~1. We recall that the sensitivity 
(\ref{signif}), {\it via} (\ref{chisq}) and its square root, 
determines the reach in $\Lambda_{\alpha\beta}$. 

%%%%%%%%%%%%%%%%%%%%%%%%%%%%%%%%%%%%%%%%%%%%%%%%%%%%%%%%%%%%%%%%%%%%%%%%
\begin{figure}[thb]
\refstepcounter{figure}
\label{Fig:Lambda-m}
\addtocounter{figure}{-1}
\begin{center}
\setlength{\unitlength}{1cm}
\begin{picture}(12.0,17.5)
\put(-1.5,+11.5){
\mbox{\epsfysize=7.2cm\epsffile{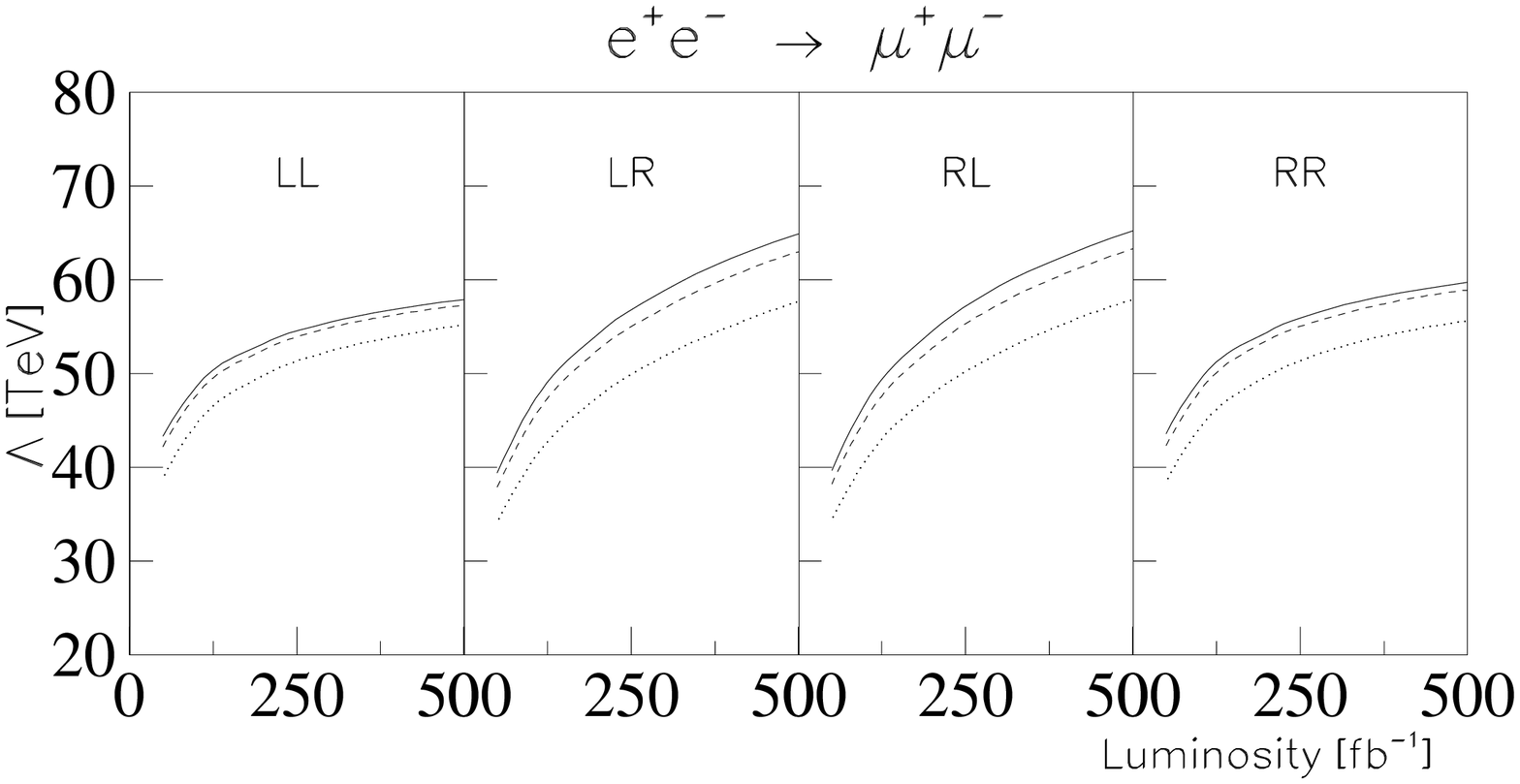}}}
\put(-1.5,5.0){
\mbox{\epsfysize=7.2cm\epsffile{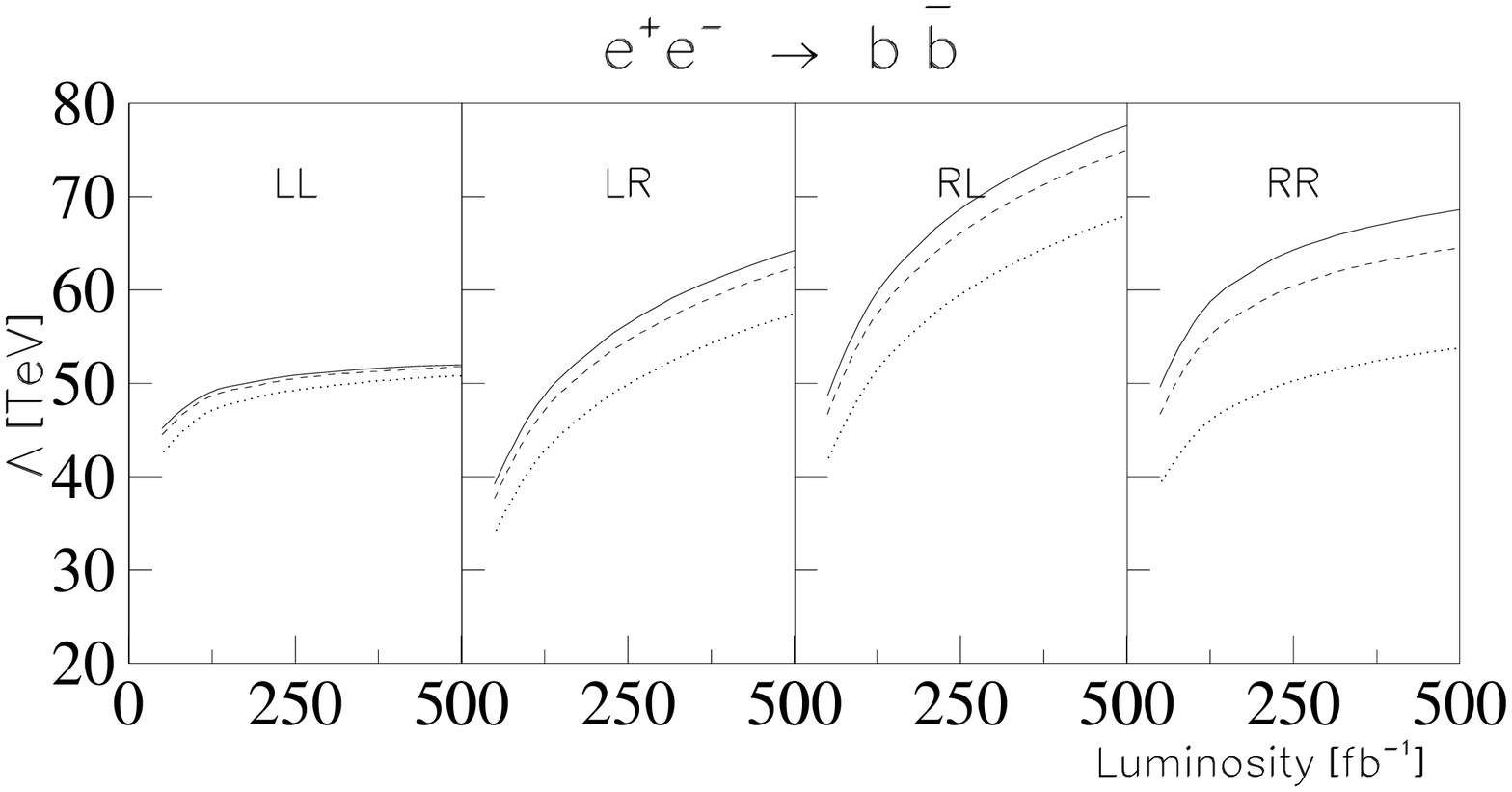}}}
\put(-1.5,-1.5){
\mbox{\epsfysize=7.2cm\epsffile{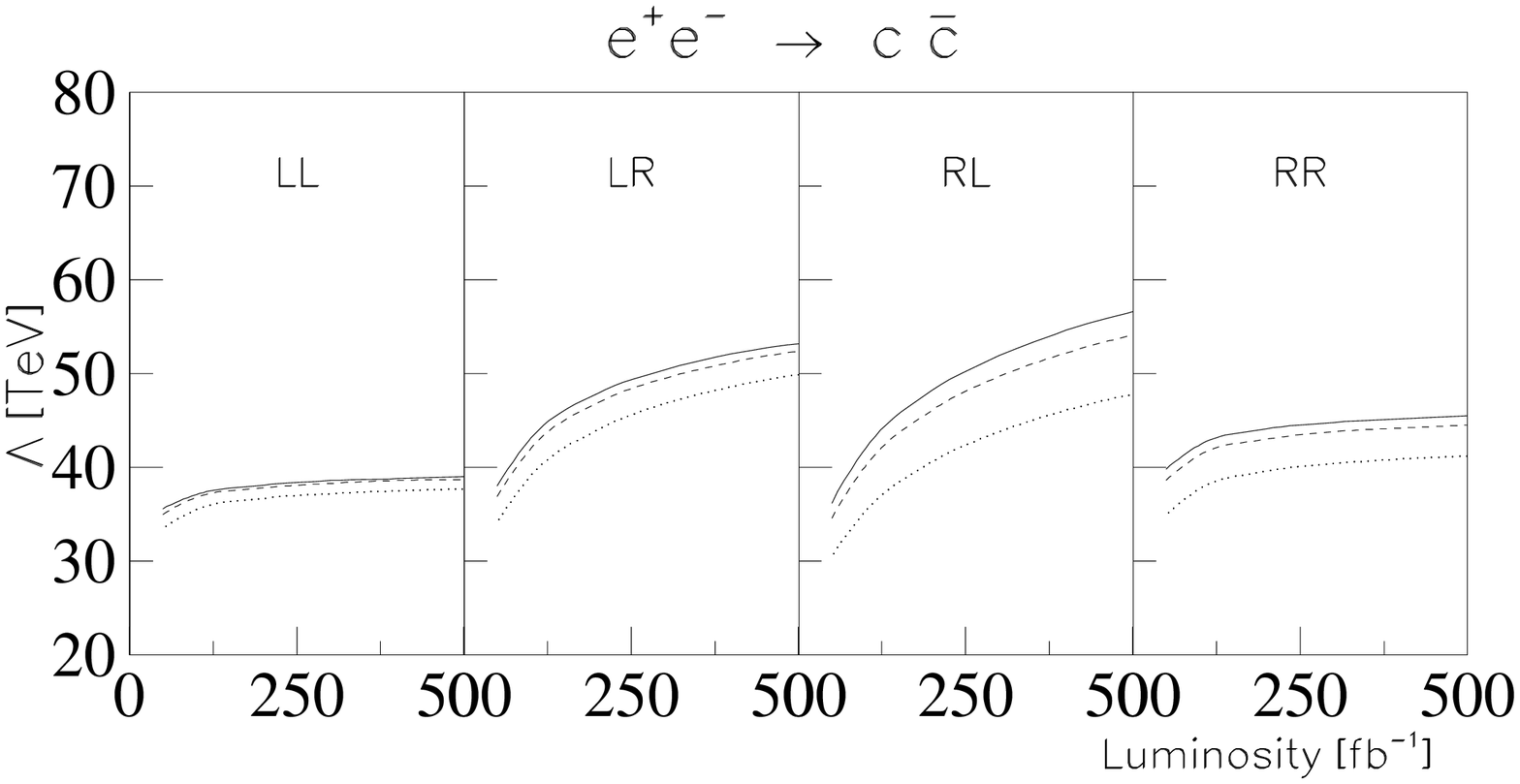}}}
\end{picture}
\vspace*{15mm}
\caption{
Reach in $\Lambda_{\alpha\beta}$ at 95\% C.L., for the proposed 
model-independent analysis, for $e^+e^-\to\mu^+\mu^-, b \bar b$ and $c \bar c$
{\it vs.} $\Lumint$. Dotted: $P_{\bar e}=0.0$;
dashed: $P_{\bar e}=0.4$; solid: $P_{\bar e}=0.6$.}
\end{center}
\end{figure}
%%%%%%%%%%%%%%%%%%%%%%%%%%%%%%%%%%%%%%%%%%%%%%%%%%%%%%%%%%%%%%%%%%%%%%%%

In the (simpler) example of polarized electrons and unpolarized 
positrons, the relative uncertainties  
$\delta\sigma_{\alpha\beta}/\sigma_{\alpha\beta}
\simeq \delta\sigma^{\rm SM}_{\alpha\beta}/\sigma^{\rm SM}_{\alpha\beta}$
compared to the case of same $P_e$, but $\delta P_e=0$,  
has been discussed for variable $\delta P_e$ in \cite{BOPP-00}. 
For the values considered here, the contribution of $\delta P_e$ to the 
overall $\delta\sigma_{\alpha\beta}$ is really negligible for the $\mu^+\mu^-$ 
final state (much less than a fraction of a \%), 
and the origin for such 
a strong suppression can be easily understood on using the SM-values
for $\sigma_1$ and $\sigma_2$ in the parameterization (\ref{Eq:uncert-P})
and (\ref{Eq:f}). 
For quarks, the contribution of $\delta P_e$ to 
$\delta\sigma_{\alpha\beta}$ turns out to be potentially larger, 
especially for the LL and RR cases of $b \bar b$ final states, but is still 
insignificant for $\delta P_e/P_e=0.5\%$ (see Fig.~2 of \cite{BOPP-00}).   

Turning to the case of both electron and positron longitudinal polarization, 
and referring to Eqs.~(\ref{s+}) and (\ref{s-}), in the chosen helicity
configuration where $P_e P_{\bar{e}}<0$, one has  $D>1$ and 
$|P_{\rm eff}|>\max(|P_e|,|P_{\bar{e}}|)$, and consequently 
an increase of the sensitivity, provided the luminosity remains the same.
However, this improvement from positron polarization is obtained up to
a maximum value of $\delta P_{\bar{e}}/P_{\bar{e}}$, above which
there would be no benefit, but, actually, a worsening of the 
sensitivity (see Figs.~3 and 4 of \cite{BOPP-00}).\footnote{Also 
this dependence can be qualitatively understood 
from (\ref{Eq:uncert-P}) and (\ref{Eq:f}).} 
This is not the case for the present input value of 
$\delta P_{\bar e}/P_{\bar e}$  
and, indeed, Fig.~1 shows a clear benefit from positron polarization 
in improving the reach on $\Lambda_{\alpha\beta}$, by about 20--40\% depending 
on the helicity configuration and the final fermion state.

%%%%%%%%%%%%%%%%%%%%%%%%%%%%%%%%%%%%%%%%%%%%%%%%%%%%%%%%%%%%%%%%%%%%%%%%
\begin{figure}[thb]
\refstepcounter{figure}
\label{Fig:Lambda-ddp}
\addtocounter{figure}{-1}
\begin{center}
\setlength{\unitlength}{1cm}
\begin{picture}(12.0,17.5)
\put(-1.5,+11.5){
\mbox{\epsfysize=7.2cm\epsffile{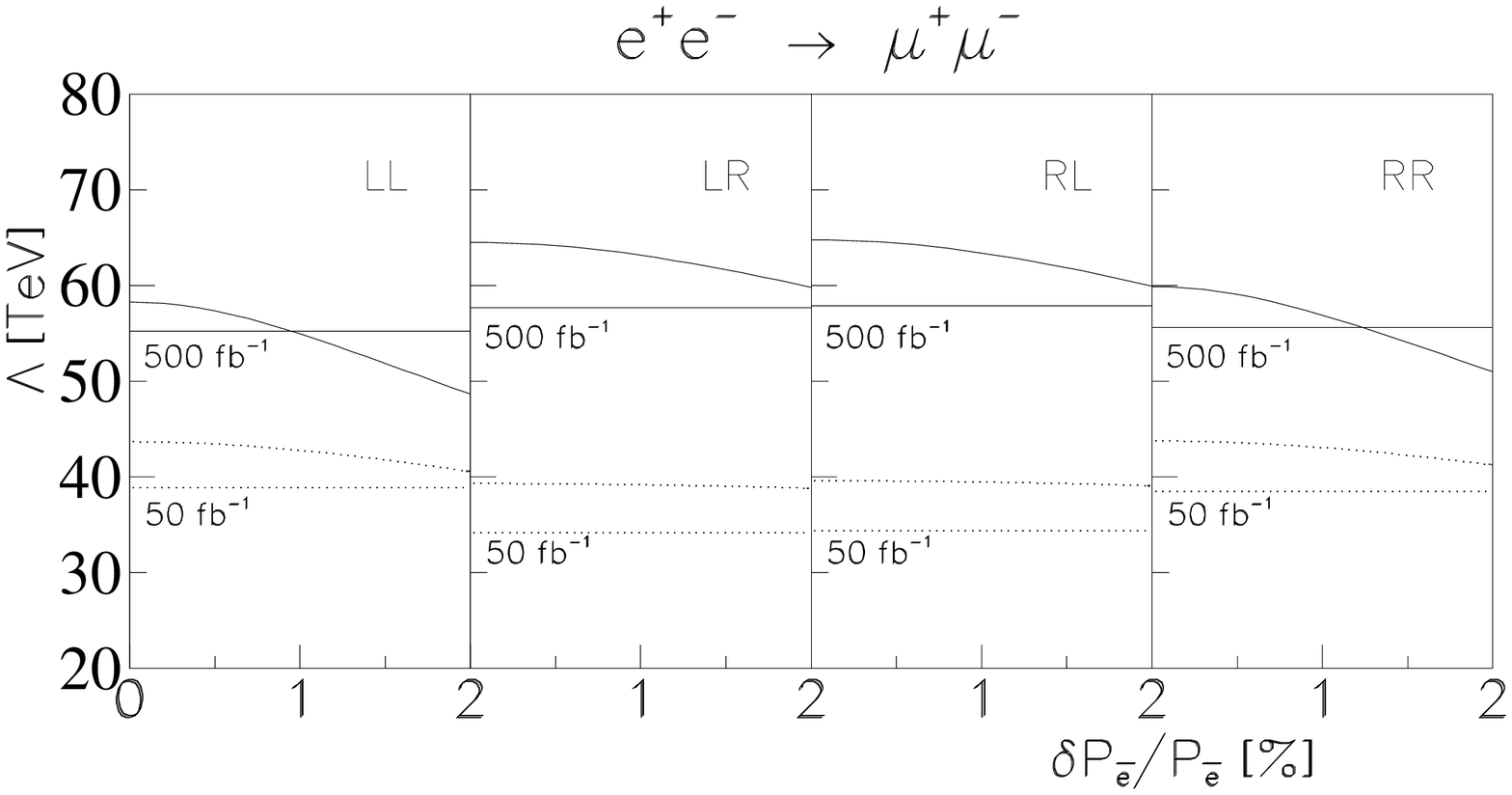}}}
\put(-1.5,5.0){
\mbox{\epsfysize=7.2cm\epsffile{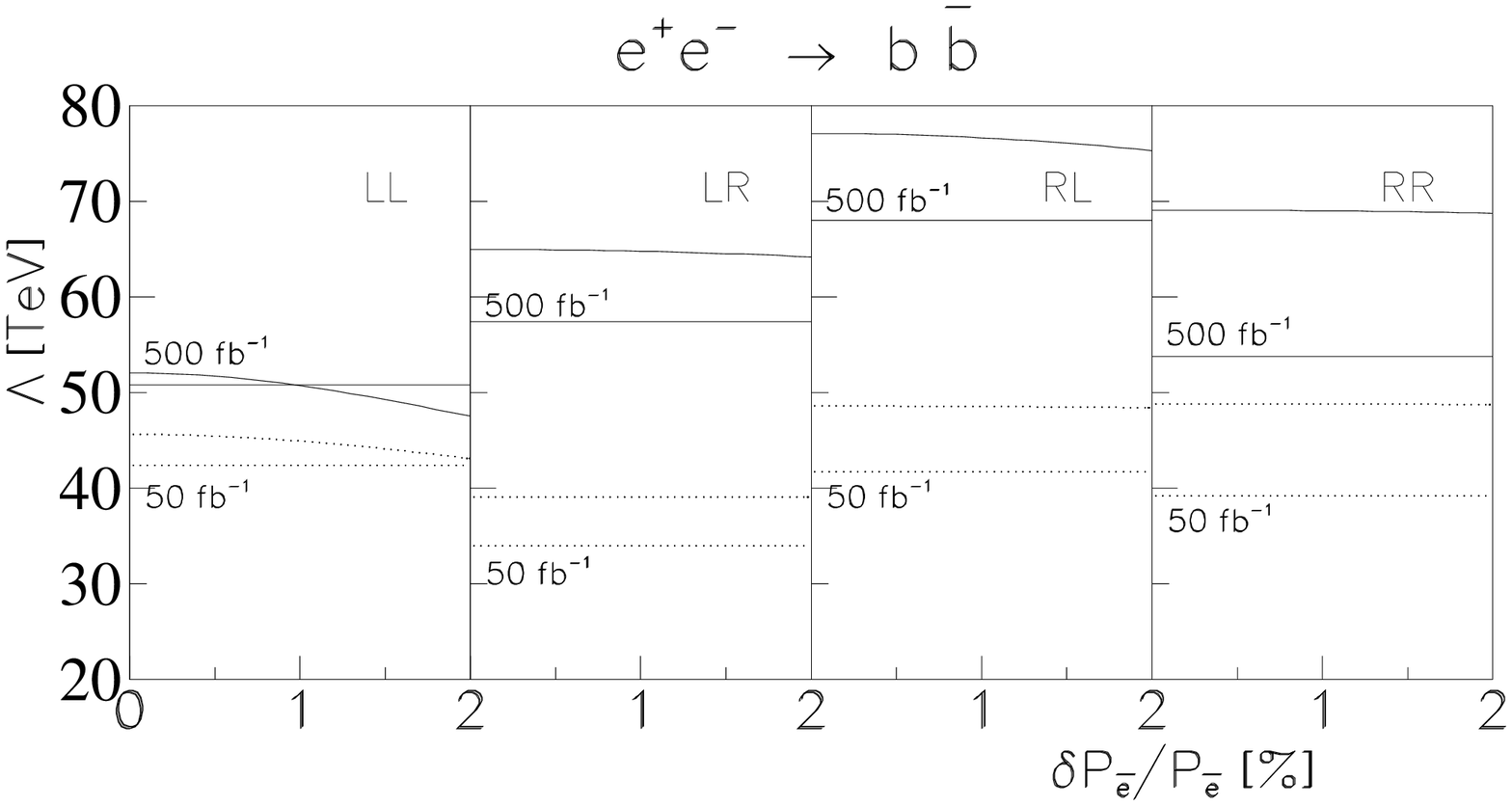}}}
\put(-1.5,-1.5){
\mbox{\epsfysize=7.2cm\epsffile{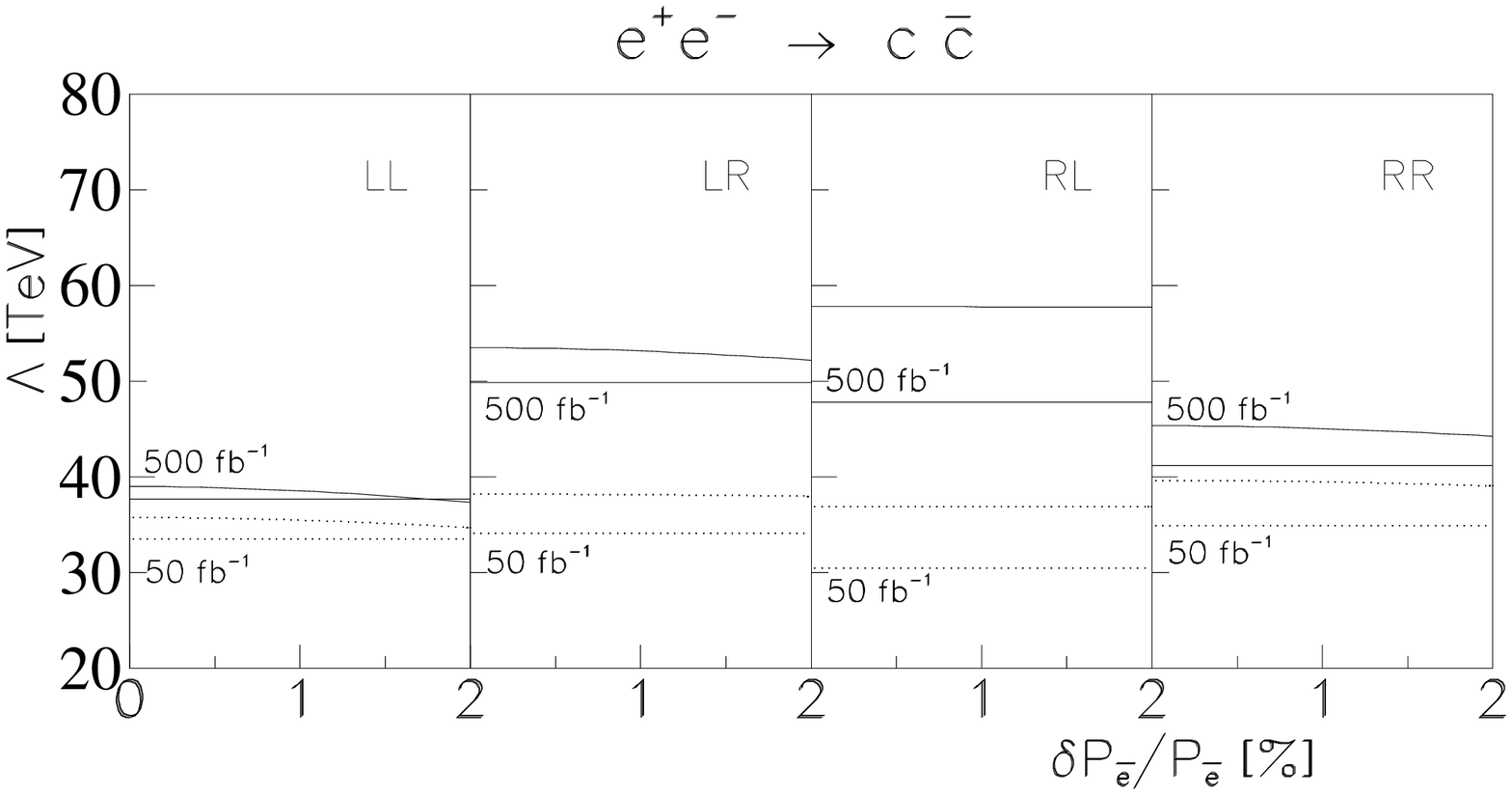}}}
\end{picture}
\vspace*{15mm}
\caption{
Reach in $\Lambda_{\alpha\beta}$ {\it vs.}\ uncertainty in positron 
polarization,
$\delta P_{\bar e}/P_{\bar e}$ for $\mu^+\mu^-$,
$b \bar b$ and $c \bar c$ final states.
Dashed: 50~$\mbox{fb}^{-1}$, solid: 500~$\mbox{fb}^{-1}$.
Horizontal lines: no positron polarization.}
\end{center}
\end{figure}
%%%%%%%%%%%%%%%%%%%%%%%%%%%%%%%%%%%%%%%%%%%%%%%%%%%%%%%%%%%%%%%%%%%%%%%%
As an indication of the influence of $\delta P_{\bar e}$, we report 
in Fig.~2 the results on $\Lambda_{\alpha\beta}$ obtained by increasing
$\delta P_{\bar e}/P_{\bar e}$ up to 2\%, with the same value
of $\delta P_e/P_e$ as in Fig.~1, and for integrated luminosity
$\Lumint=50$ and 500~$\mbox{fb}^{-1}$.
The anticipated worsening of the constraints for increasing
$\delta P_{\bar e}$ is well represented in Fig.~2.
Clearly, the point at which the benefit from positron polarization
would be lost, is determined by the relative sizes of the uncertainties
due to $\delta P_{\bar e}/P_{\bar e}$ 
and the other sources of experimental uncertainties,
in particular by the specific reference values adopted in the
parameterization of the $\delta\sigma_i$.\footnote{Of course, 
a similar discussion applies to the role of $\delta P_e$.}
On the other hand, there is some confidence that $P_e$ and $P_{\bar e}$
could be measured with the same kind of precision \cite{schuler},
so that full benefit from positron polarization should be obtained.

Fig.~1 shows that, at higher luminosity, all curves 
become less steep. This is a reflection of the fact that the statistical
uncertainty decreases with respect to the other ones,
including those due to polarization uncertainties.
Therefore, one can expect a saturation of these curves when
the integrated luminosity is such that the statistical uncertainty
becomes negligible (of course, this depends on the individual channels
and helicity combination), unless the systematic uncertainties are
diminished accordingly.
For example, for $\Lumint=10^3~\mbox{fb}^{-1}$, we would obtain 
for $P_{\bar e}=0.6$, and for $\mu^+\mu^-$ final states,
the lower bounds $\Lambda_{\rm LL}=60~\mbox{TeV}$,
$\Lambda_{\rm LR}=73~\mbox{TeV}$, $\Lambda_{\rm RL}=73~\mbox{TeV}$
and $\Lambda_{\rm RR}=62~\mbox{TeV}$.
The corresponding numbers for the hadronic channels are, respectively:
53, 72, 86 and 72~TeV for $b\bar b$; 
39, 56, 62 and 46~TeV for $c\bar c$.

In this regard, for completeness one should discuss at the same time
also the dependence of the bounds on $\Lambda_{\alpha\beta}$ from
the systematic uncertainty $\delta^{\rm sys}$ of Eq.~(\ref{delsi1}).
Basically, the effect of $\delta^{\rm sys}$ variations around 
the chosen input values in the derivation of Figs.~1 and 2
is rather small in the LR and RL cases where the statistical uncertainty 
is the dominant one for the luminosities
considered here, but can be appreciable in the LL and RR cases where
statistical and systematic uncertainties are comparable.

One can note that the bounds on $\Lambda_{\alpha\beta}$,
although derived for the most general case, where all the contact
interaction couplings of Eq.~(\ref{lagra}) simultaneously appear
as free parameters, are numerically comparable to those
obtained by allowing the presence of just one specific helicity
channel at a time \cite{Riemann-00}.
In this connection, a significant role is played by the optimization 
procedure introduced previously, i.e., the introduction of the optimal
kinematical value of $z^*$ in the definition of $\sigma_1$ and $\sigma_2$
of Eqs.~(\ref{sigma1}) and (\ref{sigma2}).
Indeed, the results on $\Lambda_{\rm LL}$ and $\Lambda_{\rm RR}$
found at such $z^*_{\rm opt}$ show a rather modest improvement
over those derived, for the same helicity combinations, from the
more conventional choice $z^*=0$ (that assumes the polarized total
cross section and forward-backward asymmetry as fundamental observables).
Conversely, the choice $z^*=z^*_{\rm opt}$ allows a dramatic improvement
of the sensitivity in the LR and RL cases,
and substantially increases the bounds on $\Lambda_{\rm LR}$
and $\Lambda_{\rm RL}$, by about 20--30\%.
For the sake of making a model-independent analysis, this improvement
certainly justifies the elaborate procedure of determining
$z^*_{\rm opt}$ from the analysis of the $z^*$ dependence
of the experimental uncertainty on $\sigma_{\alpha\beta}$
as measured {\it via} $\sigma_1$ and $\sigma_2$, prior to the application
of the $\chi^2$ procedure for the derivation of constraints on 
the $\Lambda_{\alpha\beta}$.

In conclusion, although the numerical support is based on the specific 
example worked out here on hypothetical values and assumptions
on the initial beam polarizations and the values and properties of 
the experimental uncertainties, the above considerations should hold
in general. Clearly, in practice, definite quantitative statements
should await a clarification of the realistic experimental situation,
in particular concerning the different sources, and relative roles,
of expected experimental errors.

%%%%%%%%%%%%%%%%%%%%%%%%%%%%%%%%%%%%%%%%%%%%%%%%%%%%%%%%%%%%%%%%%%%%%%%%
%\section*{Acknowledgements}
\medskip
\leftline{\bf Acknowledgements}
\par\noindent
This research has been supported by the Research Council of Norway,
and by MURST (Italian Ministry of University, Scientific Research
and Technology).
%%%%%%%%%%%%%%%%%%%%%%%%%%%%%%%%%%%%%%%%%%%%%%%%%%%%%%%%%%%%%%%%%%%%%%%%
\goodbreak

%%%%%%%%%%%%%%%%%%%%%%%%%%%%%%%%%%%%%%%%%%%%%%%%%%%%%%%%%%%%%%%%%%%%%%%%

\end{document}